\documentclass[aps,prl,reprint,amsmath,amssymb]{revtex4-1}
\usepackage{amsmath}
\usepackage{bm}
\usepackage{color}
\usepackage{graphicx}
\usepackage[pdftex,colorlinks]{hyperref}
\usepackage{times}

\let\de=\partial
\DeclareMathOperator{\tr}{tr}
\DeclareMathOperator{\sn}{sn}
\newcommand{\La}{\mathcal{L}}
\newcommand{\Ha}{\mathcal{H}}
\newcommand{\imag}{\text{i}}
\newcommand{\De}{\mathcal{D}}
\newcommand{\Pe}{\mathcal{P}}
\newcommand{\gr}[1]{\text{#1}}
\newcommand{\vek}[1]{\bm{#1}}
\allowdisplaybreaks

\begin{document}

\title{Anomaly-Induced Inhomogeneous Phase in Quark Matter without the Sign Problem}

\author{Tom\'{a}\v{s} Brauner}
\email{tomas.brauner@uis.no}
\affiliation{Department of Mathematics and Physics, University of Stavanger, 4036 Stavanger, Norway}

\author{Georgios Filios}
\email{georgios.filios@uis.no}
\affiliation{Department of Mathematics and Physics, University of Stavanger, 4036 Stavanger, Norway}

\author{Helena Kole\v{s}ov\'{a}}
\email{helena.kolesova@uis.no}
\affiliation{Department of Mathematics and Physics, University of Stavanger, 4036 Stavanger, Norway}

\begin{abstract}
We demonstrate the existence of an anomaly-induced inhomogeneous phase in a class of vector-like gauge theories without sign problem, thus disproving the long-standing conjecture that the absence of sign problem precludes spontaneous breaking of translational invariance. The presence of the phase in the two-color modification of quantum chromodynamics can be tested by an independent nonperturbative evaluation of the neutral pion decay constant as a function of external magnetic field. Our results provide a benchmark for future lattice studies of inhomogeneous phases in dense quark matter.
\end{abstract}

\maketitle


\emph{Introduction.}---Self-organization of matter into inhomogeneous patterns is ubiquitous in nature; after all, most natural materials develop crystalline order at sufficiently low temperatures. Yet, in quantum field theory, one usually assumes that the ground state of a given quantum system is uniform, unless a specific mechanism for structure formation is in place. The question under what conditions the ground state \emph{can} be nonuniform does not seem to have a satisfactory answer.

Various nonuniform phases are expected to play an important role for the thermodynamics of quark matter under extreme conditions~\cite{Buballa:2014tba}. Predictions of such phases for the phase diagram of Quantum ChromoDynamics (QCD) are, however, mostly based on model calculations neglecting order parameter fluctuations, which may be crucial for the (in)stability of the phase~\cite{Baym:1982ca,*Lee:2015bva,*Hidaka:2015xza}. Scarce attempts to study inhomogeneous states of quark matter in ab initio simulations are limited to simplified models in two spacetime dimensions~\cite{deForcrand:2006zz,*Pannullo:2019bfn} or crude approximations to the QCD functional integral~\cite{Yamamoto:2014lia}. More systematic first-principle investigation has been impeded by the notorious sign problem~\footnote{Here we \emph{define} the sign problem by the property that the determinant of the Euclidean Dirac operator of a given theory is not real and non-negative, thus precluding a straightforward application of lattice Monte Carlo techniques to the theory. The same term is sometimes used in a broader sense, depending on the particular representation of the functional integral of the theory.}.

In fact, Ref.~\cite{Splittorff:2000mm} puts forward an intriguing hypothesis, linking the appearance of nonuniform states in the phase diagram of a vector-like gauge theory (hereafter referred to as ``QCD-like theories'') to the very presence of the sign problem in the theory. If true, this would provide a rare example of a no-go theorem for spontaneous breaking of a spacetime symmetry.

In this Letter, we disprove this conjecture. We demonstrate that a class of QCD-like theories free of the sign problem features the nonuniform Chiral Soliton Lattice (CSL) phase~\footnote{In the limit of vanishing quark mass, this state, sometimes referred to as a ``meson supercurrent'', was discussed in Ref.~\cite{Bergman:2008qv,*Thompson:2008qw,*Rebhan:2008ur}.}, previously shown to exist in QCD itself~\cite{Son:2007ny,Brauner:2016pko}. This state is a remarkable manifestation of the chiral anomaly, and requires subjecting quark matter to a magnetic field or to global rotation~\cite{Huang:2017pqe}; see also Ref.~\cite{Fukushima:2018ohd,*Kawaguchi:2018fpi} for closely related recent work.

To the best of our knowledge, this is the first time that existence of an inhomogeneous phase in gauge theories amenable to direct lattice Monte Carlo simulation has been shown. Our results can thus serve as a benchmark for future ab initio studies of nonuniform phases in dense quark matter, or elsewhere.


\emph{Absence of sign problem.}---We consider the class of QCD-like theories where quarks transform in a (pseudo)real representation of the gauge group~\cite{Kogut:1999iv,*Kogut:2000ek}, restricting ourselves for simplicity to two degenerate quark flavors $u,d$ with the common current mass $m$. Let us denote the Euclidean Dirac operator for a single quark flavor as $\De_i\equiv\gamma_\mu D_{i\mu}+m-\mu\gamma_0$, where $i=u,d$. Here $\mu$ is the quark number chemical potential and the Hermitian Euclidean Dirac matrices $\gamma_\mu$ satisfy the charge conjugation property $(C\gamma_5)\gamma_\mu(C\gamma_5)^{-1}=\gamma_\mu^*$. Finally, the covariant derivative is defined by $D_{i\mu}\equiv\de_\mu-\imag T_aA^a_\mu-\imag q_i A^Q_\mu$, where $A^a_\mu$ are the gluon fields, $q_i$ the quark electric charge, and $A^Q_\mu$ represents a background electromagnetic field.

In (pseudo)real QCD-like theories, the color generators $T_a$ satisfy by assumption $T_a^*=-\Pe T_a\Pe^{-1}$. Without loss of generality, the matrix $\Pe$ can be assumed unitary and symmetric for real quarks, and unitary and antisymmetric for pseudoreal quarks~\cite{Georgi:1982jb}. For instance, for a theory with the $\gr{SU(2)}$ gauge group and fundamental quarks (``two-color QCD''), $\Pe$ is given by a Pauli matrix in the color space, $\Pe=\sigma_2$.

Provided the electric charges of the $u$- and $d$-quarks satisfy $q_u=-q_d$, their respective Dirac operators are related by
\begin{equation}
(KC\gamma_5\Pe)\De_u=\De_d(KC\gamma_5\Pe),
\end{equation}
where $K$ is the operator of complex conjugation. This establishes an antiunitary mapping between the eigenvectors of $\De_u$ and $\De_d$~\cite{Hands:2000ei}. Hence the determinant of the Dirac operator of the theory is real and non-negative, $\det\De=\det\De_u\det\De_d=\det\De_u\det\De_u^*\geq0$. To conclude, (pseudo)real QCD-like theories with two quark flavors of equal masses are free of the sign problem in presence of both quark number chemical potential and external electromagnetic field, as long as the two quark flavors have opposite electric charges.


\emph{(Pseudo)real theories in magnetic field.}---We shall further assume that the color gauge group and its (pseudo)real quark representation are chosen so that the theory has a confining, chiral-symmetry-breaking vacuum just like QCD. The low-energy physics of the theory is then dominated by the pseudo-Nambu-Goldstone bosons of its flavor symmetry.

In the limit of zero quark mass (``chiral limit''), (pseudo)real QCD-like theories with $N$ quark flavors feature an enhanced $\mathcal G=\gr{SU}(2N)$ flavor symmetry, which includes as its generators both the electric charge $Q$ and the baryon number $B$. The chiral condensate in the ground state breaks this spontaneously to $\mathcal H=\gr{SO}(2N)$ in real theories, and to $\mathcal H=\gr{Sp}(2N)$ in pseudoreal theories, resulting in $2N^2+N-1$ pseudo-Nambu-Goldstone bosons in the real case, and $2N^2-N-1$ ones in the pseudoreal case. Of these, $N^2-1$ are pseudoscalar mesons (``pions''), while the remaining modes, absent in QCD, are scalar diquarks. All the modes share the same mass, $m_\pi$, assuming equal current masses of all quark flavors~\cite{Kogut:1999iv,Kogut:2000ek}.

For $N=2$ and with the choice of charges $q_u=-q_d\neq0$, a uniform external magnetic field reduces the symmetry to
\begin{equation}
\begin{split}
\mathcal G_Q&=\gr{SU}(2)\times\gr{SU}(2)\times\gr{U}(1)_Q,\\
\mathcal H_Q&=\gr{SU}(2)_\text{diag}\times\gr{U}(1)_Q.
\end{split}
\label{flavor_symmetry}
\end{equation}
Note that $\mathcal G_Q$ is not the usual chiral symmetry of two-flavor QCD: the baryon number $B$ is included as a generator of the ``vector'' subgroup $\gr{SU}(2)_\text{diag}$. The number of electrically neutral light degrees of freedom, given by the dimension of the coset space $\mathcal G_Q/\mathcal H_Q$, is three in both real and pseudoreal theories, including the neutral pion $\pi^0$ and an electrically neutral diquark-antidiquark pair $d^0,\bar d^0$.


\begin{figure}
\includegraphics[width=\columnwidth]{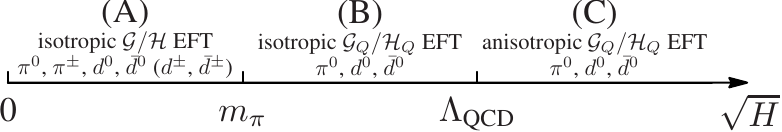}
\caption{Different regimes of the EFT and the corresponding light degrees of freedom, depending on the strength of the magnetic field. For $\sqrt H\ll\Lambda_\text{QCD}$, the magnetic field can be treated as a perturbation of the ground state of the QCD-like theory. For $\sqrt H\gtrsim\Lambda_\text{QCD}$, the ground state is strongly affected by the field and the low-energy EFT becomes anisotropic. For $\sqrt H\gg m_\pi$, the charged degrees of freedom become heavy and decouple from the EFT. (The light charged diquarks $d^\pm$, $\bar d^\pm$ are only present in real QCD-like theories.)}
\label{Fig:axis}
\end{figure}

\emph{Low-energy effective theory.}---In magnetic fields $H\gg m_\pi^2$, charged (pseudo)scalars become heavy due to Landau level quantization, and the low-energy physics will be dominated by the electrically neutral modes. In this regime, which we will from now on assume, a low-energy effective theory (EFT) based on the coset space $\mathcal G_Q/\mathcal H_Q$ can be constructed by using its isomorphism with that of two-flavor QCD. 

Magnetic fields around the characteristic scale of the theory, $\Lambda_\text{QCD}$, or stronger, will distort the ground state and make the low-energy EFT anisotropic, breaking the Lorentz group $\gr{SO}(3,1)$ down to $\gr{SO}(1,1)\times\gr{SO}(2)$. See Fig.~\ref{Fig:axis} for a sketch of the different regimes of the EFT. We aim at finding an effective action for the strong-field regime (C). This can be done by contracting Lorentz indices with projections, $g_{\parallel\mu\nu}$ and $g_{\perp\mu\nu}$, of the Minkowski metric to the two-dimensional subspaces left intact by the magnetic field. Our EFT will also be valid in the moderate-field regime (B), albeit with a reduced predictive power due to the use of a lower spacetime symmetry.

Taking finally into account the discrete symmetries $C$, $P$, $T$, the leading-order effective Lagrangian is given by~\cite{Miransky:2002rp,*Miransky:2015ava}
\begin{equation}
\begin{split}
\La_\text{eff}=\frac{f_\pi^2}4\bigl[&(g^{\mu\nu}_\parallel+v^2g^{\mu\nu}_\perp)\tr(D_\mu\Sigma D_\nu\Sigma^{-1})\\
&+m_\pi^2\tr(\Sigma+\Sigma^{-1})\bigr]+\La_\text{WZ}.
\end{split}
\label{lagrangian}
\end{equation}
Here $v$ is a velocity parameter and $f_\pi$, $m_\pi$ are the pion decay constant and mass, respectively. All the parameters $f_\pi$, $m_\pi$, $v$ are given by a priori unknown functions of the magnetic field. The $2\times2$ unimodular unitary matrix field $\Sigma$ contains the three electrically neutral degrees of freedom. The covariant derivative $D_\mu\Sigma$, specified below in Eq.~\eqref{covder}, introduces the coupling of diquarks to baryon number chemical potential.

The $\La_\text{WZ}$ piece in Eq.~\eqref{lagrangian}, known as the Wess-Zumino (WZ) term~\cite{Wess:1971yu,*Witten:1983tw}, is the contribution of the chiral anomaly. This can also be found following the analogy with two-flavor QCD, by swapping the roles of electric charge and baryon number in the result given in Ref.~\cite{Son:2007ny}. For $q_u=-q_d\neq0$, it reads
\begin{align}
\label{WZterm}
\La_\text{WZ}=&-\frac C6\epsilon^{\mu\nu\alpha\beta}A^Q_\mu\tr(\de_\nu\Sigma\de_\alpha\Sigma^{-1}\de_\beta\Sigma\Sigma^{-1})\\
\notag
&+\frac{\imag bC}4\epsilon^{\mu\nu\alpha\beta}F^Q_{\mu\nu}A^B_\alpha\tr[\tau_3(\de_\beta\Sigma\Sigma^{-1}-\de_\beta\Sigma^{-1}\Sigma)],
\end{align}
$b$ being the baryon number of a single quark, $F^Q_{\mu\nu}$ the electromagnetic field strength, $A^B_\mu$ an external gauge potential that couples to the baryon number current, and $\tau_3$ a Pauli matrix in the flavor space. The overall normalization of the WZ term is not determined by symmetry, but can be fixed by matching the EFT to the underlying QCD-like theory. One then finds~\cite{Brauner:2019aid}
\begin{equation}
C=\frac{d}{8\pi^2}(q_u-q_d),
\label{Cdef}
\end{equation}
where $d$ is the dimension of the representation of the color gauge group that a single quark flavor transforms in.

A few remarks are in order here. First, while we exploited the analogy with two-flavor QCD, the form of the EFT can as well be obtained by first constructing the EFT for the full coset space $\mathcal G/\mathcal H$ and then discarding all charged degrees of freedom. This requires the knowledge of the gauged WZ term in (pseudo)real QCD-like theories though~\cite{Duan:2000dy,*Brauner:2018zwr}.

Second, the assumption that the external magnetic field satisfies $H\gg m_\pi^2$, that is lies in the regime (B) or (C), is of little practical limitation. We shall see that interesting physics occurs above certain critical value of the field, which is safely above the (A) regime for light quarks.

Finally, in contrast to the chiral perturbation theory of QCD (see Ref.~\cite{Ecker:1994gg,*Pich:1995bw,*Scherer:2002tk} for a review), the WZ term~\eqref{WZterm} contributes to the leading order of the EFT. This is due to a modified power counting, whereby the baryon gauge field $A^B_\mu$ counts, just like all derivatives, as $\mathcal O(p^1)$, whereas the magnetic field $H$ counts as $\mathcal O(p^0)$. The latter is required for consistency of the EFT in the (C) regime, and makes the coefficients $f_\pi$, $m_\pi$, $v$ functions of $H$. The resulting EFT may be valid for arbitrarily strong fields as long as the ground state breaks the flavor symmetry by the formation of the chiral condensate, which is supported by the magnetic catalysis phenomenon~\cite{Miransky:2002rp,*Miransky:2015ava}. Note that, in contrast, in the (A) regime $f_\pi$, $m_\pi$ are true constants (and $v=1$) and the dependence of the EFT on $H$ is fully fixed by electromagnetic gauge invariance.

In the rest of this Letter, we analyze the ground state of the EFT at nonzero baryon number chemical potential, $\mu_B$, and magnetic field $H$. Details of excitation spectrum in the various phases in the phase diagram are reported elsewhere~\cite{Brauner:2019aid}. We will fix without loss of generality $q_u=-q_d=1/2$ and $b=1/2$; any other choice can be absorbed into a redefinition of $H$ and $\mu_B$, respectively.

To bring the EFT into a form more suitable for the analysis, we map the matrix $\Sigma$ on a unit four-vector, $\Sigma\equiv n_0+\imag\vec n\cdot\vec\tau$, where $\vec\tau$ are the Pauli matrices and $n_0^2+\vec n^2=1$. The nontrivial components of the covariant derivative $D_\mu\Sigma$ then read
\begin{equation}
D_0n_1\equiv\de_0n_1-\mu_Bn_2,\quad
D_0n_2\equiv\de_0n_2+\mu_Bn_1.
\label{covder}
\end{equation}
The effective Lagrangian~\eqref{lagrangian} now boils down to
\begin{align}
\label{Leff}
\La_\text{eff}={}&\frac{f_\pi^2}2(g^{\mu\nu}_\parallel+v^2g^{\mu\nu}_\perp)(\de_\mu n_0\de_\nu n_0+\de_\mu\vec n\cdot\de_\nu\vec n)\\
\notag
&+f_\pi^2\mu_B(n_1\de_0n_2-n_2\de_0n_1)+\frac{f_\pi^2}2\mu_B^2(n_1^2+n_2^2)\\
\notag
&+f_\pi^2m_\pi^2n_0+CH\mu_B(n_0\de_zn_3-n_3\de_zn_0)+\dotsb,
\end{align}
where we oriented the magnetic field along the $z$-axis and set $A^B_\mu=(\mu_B,\vek0)$. The ellipsis indicates terms with three derivatives, coming from the first line of Eq.~\eqref{WZterm}; being linear in time derivatives, they do not contribute to the Hamiltonian and thus do not affect the structure of the ground state~\footnote{This argument requires a gauge where $A^Q_0=0$. The final phase diagram must, of course, be independent of the gauge choice.}.


\emph{Chiral limit.}---The ground state is easy to determine by a direct minimization of the Hamiltonian in the chiral limit, $m_\pi\to0$. For any nonzero $\mu_B$, there turn out to be two phases. For $CH<f_\pi^2$, the ground state is $\langle n_0\rangle=\langle n_3\rangle=0$ and $\langle n_1\rangle^2+\langle n_2\rangle^2=1$. This describes a Bose-Einstein condensate (BEC) of diquarks, which appears in the phase diagram of (pseudo)real QCD-like theories generally for $\mu_B>m_\pi$~\cite{Kogut:1999iv,*Kogut:2000ek}.

For $CH>f_\pi^2$, on the other hand, the ground state features a spatially dependent chiral condensate and neutral pion condensate, but no diquark condensate: $\langle n_1\rangle=\langle n_2\rangle=0$ and
\begin{equation}
\langle n_0\rangle=\cos\frac{CH\mu_Bz}{f_\pi^2},\qquad
\langle n_3\rangle=\sin\frac{CH\mu_Bz}{f_\pi^2},
\label{CDW}
\end{equation}
up to an arbitrary translation of the $z$-coordinate. This corresponds to the CSL state in the chiral limit~\cite{Brauner:2016pko,Bergman:2008qv,*Thompson:2008qw,*Rebhan:2008ur}. By Eq.~\eqref{Cdef}, the CSL state appears in the phase diagram for
\begin{equation}
H>H_\text{cr}=\frac{8\pi^2f_\pi^2}{d}.
\label{Bcrit}
\end{equation}
The competition of the CSL and BEC phases is in a stark contrast to QCD, where in the chiral limit, the CSL state is triggered by arbitrarily weak magnetic fields. In (pseudo)real QCD-like theories, a nonzero critical field is required to overcome the energy gain of diquark BEC.


\emph{Full phase diagram.}---To get insight into the phase diagram away from the chiral limit, it is convenient to parameterize the unit four-vector variable in terms of three spherical angles,
\begin{equation}
\begin{aligned}
n_0&=\cos\theta\cos\phi,&\qquad
n_3&=\cos\theta\sin\phi,\\
n_1&=\sin\theta\cos\alpha,&\qquad
n_2&=\sin\theta\sin\alpha.
\end{aligned}
\label{angle}
\end{equation}
We also introduce dimensionless variables that allow us to scale out trivial dependence of observables on $f_\pi$ and $m_\pi$,
\begin{equation}
\bar x^\mu\equiv m_\pi x^\mu,\qquad
x\equiv\frac{\mu_B}{m_\pi},\qquad
\bar H\equiv\frac{CH}{f_\pi^2}.
\end{equation}
It is easy to see that the ground state has to be independent of time and the transverse coordinates. The task to find the ground state thus reduces to that of minimizing (the spatial average of) the one-dimensional effective Hamiltonian
\begin{align}
\label{HamDimless}
\frac{\Ha_\text{eff}}{f_\pi^2m_\pi^2}={}&\frac12\left[(\theta')^2+(\phi')^2\cos^2\theta+(\alpha')^2\sin^2\theta\right]\\
\notag
&-\frac{x^2}2\sin^2\theta-\cos\theta\cos\phi-x\bar H\phi'\cos^2\theta+1,
\end{align}
where the primes denotes derivatives with respect to $\bar z$~\footnote{The very last $+1$ term was added to ensure that the energy of the trivial vacuum is zero.}. The ground state is realized by constant $\alpha$, hence we are dealing with a one-dimensional system of two variables $\theta$ and $\phi$.

Restricting first to uniform field configurations, it readily follows that there are two candidate states: the trivial vacuum with $\langle\theta\rangle=\langle\phi\rangle=0$ and $\bar\Ha_\text{vac}\equiv\Ha_\text{eff}/(f_\pi^2m_\pi^2)=0$, and, for $x\geq1$, the diquark BEC state with
\begin{equation}
\cos\langle\theta\rangle=\frac1{x^2},\quad
\langle\phi\rangle=0,\quad
\bar\Ha_\text{BEC}=-\frac12\left(x-\frac1x\right)^2.
\end{equation}

The nonuniform CSL state, found in Ref.~\cite{Brauner:2016pko}, can be embedded into the present EFT for (pseudo)real QCD-like theories by setting $\langle\theta\rangle=0$. It satisfies
\begin{equation}
\cos\frac{\langle\phi(\bar z)\rangle}2=\sn\left(\frac{\bar z}k,k\right),
\end{equation}
where $\sn$ is one of Jacobi's elliptic functions and $k$ the corresponding elliptic modulus. This describes a periodic soliton with lattice spacing $\ell=2kK(k)m_\pi^{-1}$, $K(k)$ being the complete elliptic integral of the first kind. The optimum value of $k$ is found by minimization of the average energy density carried by the soliton, and fulfills the condition
\begin{equation}
\frac{E(k)}k=\frac{\pi x\bar H}4,
\label{CSLvacuum}
\end{equation}
$E(k)$ being the complete elliptic integral of the second kind. The energy of the CSL state can be cast implicitly as
\begin{equation}
\bar\Ha_\text{CSL}=2\left(1-\frac1{k^2}\right).
\end{equation}
Owing to $0\leq k\leq1$, this state always has a lower energy than the trivial vacuum, but only exists for $\bar H\geq4/(\pi x)$.

\begin{figure}
\includegraphics[width=\columnwidth]{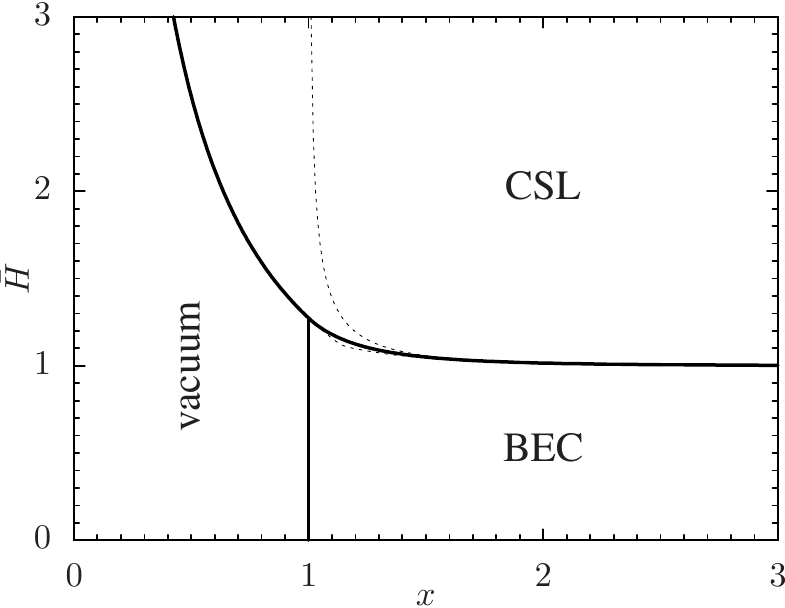}
\caption{Tentative phase diagram in the $\bar H\equiv CH/[f_\pi(H)]^2$ and $x\equiv\mu_B/m_\pi(H)$ variables at zero temperature. The solid lines denote phase transitions. The dashed lines are the spinodal curves of the first-order transition between the BEC and CSL phases, obtained from the analysis of the excitation spectrum~\cite{Brauner:2019aid}.}
\label{Fig:phase_diagram}
\end{figure}

Comparing the energies of the BEC and CSL states leads to the phase diagram in Fig.~\ref{Fig:phase_diagram}. While this was found with simple ansatz stationary states, we have strong, analytical and numerical, evidence based on a variational treatment of the Hamiltonian~\eqref{HamDimless} that no other state of even lower energy exists~\cite{Brauner:2019aid}. We can also conclude rigorously that the ground state in the ``CSL'' region in Fig.~\ref{Fig:phase_diagram}, whatever it is, has to be nonuniform.

\emph{Discussion and summary.}---According to Fig.~\ref{Fig:phase_diagram}, fields with $\bar H>1$ are required to generate a spatially modulated ground state. While our EFT is in principle valid for arbitrarily strong fields, it is nevertheless not clear without detailed knowledge of the function $f_\pi(H)$ whether $\bar H>1$ can be satisfied for any physical values of $H$.

Rewriting this condition as $H/(4\pi f_\pi)^2>1/(2d)$, and recalling that the loop factor $4\pi f_\pi$ controls the derivative expansion of the EFT~\cite{Manohar:1983md}, we expect that the critical field for the formation of CSL can be reached in theories with sufficiently large $d$. Indeed, a one-loop calculation within the chiral perturbation theory of QCD gives, in the chiral limit~\cite{Shushpanov:1997sf,*Agasian:2001ym},
\begin{equation}
[4\pi f_\pi(H)]^2=[4\pi f_\pi(0)]^2+2H\log 2+\mathcal O(H^2).
\label{SU2a}
\end{equation}
The same result applies to all \emph{pseudoreal} QCD-like theories, since the $H$-dependent correction to $f_\pi$ comes from a charged pion loop, and is thus insensitive to the presence of electrically neutral diquarks. By Eq.~\eqref{SU2a}, the one-loop correction to the critical magnetic field $H_\text{cr}$ will be suppressed by $1/d$, and thus negligible for large $d$. Sufficient accuracy is then achieved by treating $f_\pi$ as an $H$-independent constant.

For theories with small $d$ such as two-color QCD where $d=2$, the $H$-dependence of $f_\pi$ \emph{has to} be taken into account to see whether a field for which $\bar H>1$ exists. Here Eq.~\eqref{SU2a} alone, implying a one-loop correction to $H_\text{cr}$ of about $50\%$ at $d=2$, is not conclusive enough. The function $f_\pi(H)$ can be further constrained by utilizing the results of Ref.~\cite{Miransky:2002rp,*Miransky:2015ava}, giving its strong-field asymptotic behavior in two-color QCD,
\begin{equation}
4\pi^2[f_\pi(H)]^2=H+\dotsb\quad\text{for }H\to\infty.
\label{SU2b}
\end{equation}
Equations~\eqref{SU2a} and~\eqref{SU2b} fix the asymptotics of $f_\pi(H)$ in the $H\to0$ and $H\to\infty$ limits. Both are clearly \emph{consistent} with the existence of the CSL phase in two-color QCD: the latter implies that $\bar H=1+\dotsb$ for $H\to\infty$. However, to make a firm conclusion, we would need to know more about the behavior of $f_\pi(H)$ between the two limits, or at least the sign of the correction to the leading term in Eq.~\eqref{SU2b}. Such input does not seem to be available at the moment, as existing results (see e.g.~Ref.~\cite{Simonov:2015xta,*Avancini:2016fgq,*GomezDumm:2017jij}) are not conclusive enough for our purposes.

To summarize, we have constructed a class of counterexamples to the conjecture that in vector-like gauge theories, positivity of the determinant of the Dirac operator (i.e.~absence of the sign problem) implies absence of inhomogeneous phases in the phase diagram~\cite{Splittorff:2000mm}. The nonuniform order is realized by a topological crystalline condensate of neutral pions, and requires a sufficiently strong background magnetic field. The conjecture might still hold under more restrictive assumptions, for instance when full rotational invariance is imposed.

Our analysis utilizes low-energy EFT and is thereby model-independent. Hence, Fig.~\ref{Fig:phase_diagram} represents the mapping of the true phase diagram of (pseudo)real QCD-like theories to the space of the dimensionless variables $\bar H$ and $x$. Our results are not limited to weak magnetic fields. Consistency of the derivative expansion requires moderate chemical potentials though: using Eq.~\eqref{CDW} to estimate the gradients involved in the CSL state leads to the bound $\bar H\mu_B\ll4\pi f_\pi$, or $\bar Hx\ll4\pi f_\pi/m_\pi$.

For theories with a large enough gauge group and its representation on the quark fields, an inhomogeneous phase can be demonstrably realized with moderate magnetic fields controlled by the derivative expansion of the EFT. In the simplest and most well studied QCD-like theory---two-color QCD---the question of the existence of a nonuniform phase remains open. Assuming that our EFT remains valid in strong magnetic fields, that is the ground state at zero chemical potential carries a chiral condensate~\cite{Miransky:2015ava}, this question can, however, be answered by an independent nonperturbative evaluation of the neutral pion decay constant as a function of magnetic field.


\emph{Acknowledgments.}---We are indebted to Naoki Yamamoto for collaboration preceding the present project and for insightful comments. The study of the CSL phase in theories without sign problem was inspired by a discussion with Hiromichi Nishimura. Last but not least, we are also grateful to Thomas Cohen, Gergely Endr\H odi, Philippe de Forcrand, Simon Hands, Carlos Hoyos, Aleksi Kurkela, Eugenio Meg\'{\i}as, Andreas Schmitt and Igor Shovkovy for fruitful discussions and for asking questions that sharpened our understanding of the subject. This work has been supported by a ToppForsk-UiS grant no.~PR-10614.


\bibliography{references}


\end{document}